\documentstyle[prb,aps]{revtex}
\tightenlines

\begin{document}
\title{Phase diagram of YBa$_2$Cu$_3$O$_{7-y}$ at T$<$T$_c$ based on Cu(2)
transverse nuclear relaxation}
\author{M.V.Eremin, Yu.A.Sakhratov, A.V.Savinkov, A.V.Dooglav, I.R.Mukhamedshin,
A.V.Egorov}
\address{Kazan State University, 420008 Kazan, Russia}
\date{\today }
\maketitle

\begin{abstract}
Two maxima in transverse relaxation rate of Cu(2) nuclei in YBa$_2$Cu$_3$O$%
_{7-y}$ are observed, at T =35 K and T = 47 K. Comparison of the $^{63}$%
Cu(2) and $^{65}$Cu(2) rates at T = 47 K indicates the magnetic character of
relaxation. The enhancement at T = 47 K of fluctuating local magnetic fields
perpendicular to the CuO$_2$ planes is connected with the critical
fluctuations of orbital currents. Maximum at T = 35 K is connected with the
appearance of inhomogeneous supeconducting phase. Together with data
published to date, our experimental results allow to suggest a qualitatively
new phase diagram of the superconducting phase. \\ \\PACS: 74.25.Nf, 74.72.Bk,
74.25.Dw, 76.60.Gv
\end{abstract}

%\draft

\twocolumn

Anomalous behavior of copper nuclear transverse relaxation rate in CuO$_2$
planes of YBa$_2$Cu$_3$O$_{7-y}$ superconductors had been discovered at the
beginning of the HTSC history \cite{r1,r2,r3,r4,r5}. The sharp increase of
the transverse relaxation rate at T = 35 K occurring in the superconducting
samples with the doping level close to optimal is considered by the authors
of \cite{r2,r6,r7} as an indication of the phase transition. Ohkawa\cite{r6}
interpreted this transition as the second superconducting transition in the
copper bilayer of YBa$_2$Cu$_3$O$_{7-y}$. Kr\"amer and Mehring\cite{r7}
connect it with the transition of YBa$_2$Cu$_3$O$_{7-y}$ to the state with
the charge density waves (CDW). As the authors of\cite{r7} have pointed out,
this interpretation needs to be checked on the samples with another oxygen
doping level. It was critisized recently in\cite{r8}, and new measurements
for a slightly overdoped YBa$_2$Cu$_3$O$_{7-y}$ samples with T$_c$=89.5 K
had been presented with the well defined Cu NQR linewidth maximum not at 35
K, but at 60 K.

Our studies of Cu(2) NQR in a series of YBa$_2$Cu$_3$O$_{7-y}$ samples
confirm the existence of a transverse relaxation maximum at T = 35 K. The
new result of our study is the discovery of one maximum more, at T = 47 K.
Moreover, a maximum of this kind appeared to be reported already in few
papers\cite{r5,r9} but no proper attention had been payed to it. However,
the existence of two maxima in some cases might indicate, to our mind, in
favour of two phase transitions below T$_c$.

Specifications we suggest for the recently discussed in \cite{r10} schematic
diagram of coexistence of phases in the superconducting state near optimal
doping is shown in Fig.1. In addition to the phase of spontaneous currents
or, in other words, phase of charge density waves with the imaginary order
parameter ({\it id}-CDW) suggested earlier\cite{r11,r12,r13}, we suppose the
existence of at least one more phase in the superconducting state, namely, a
phase of the Cooper pairs with a non-zero total momentum or, in other words,
an inhomogeneous component of the superconducting {\it d}-phase ({\it d}%
-ISC). The possibility of co-existence of few phases in underdoped cuprates
is connected with a Peierls instability of a quasi-2D system and was
discussed in\cite{r11,r12,r13}. The scenario of competition between {\it id}%
-CDW and {\it d}-type superconducting ({\it d}-SC) phases explains fairly
well the non-monotonous behavior of the effective gap near T$_c$\cite{r14}.
It was used recently for explanation of neutron scattering data\cite{r15}
and anomalous behavior of spin-lattice relaxation in superconducting
cuprates at low temperatures\cite{r16}.

The dashed line in Fig.1, just as in\cite{r10}, separates the {\it id}-CDW
phase, the continuous dome-like line is a boundary of the {\it d}-SC phase.
The behavior of the phase boundary depicted by the dashed line in Fig.1
differs from that published in\cite{r10} and is obtained according to the
calculations of\cite{r11,r12}. The existence of the phase boundary depicted
by a dash-dotted line we assume according to the transverse nuclear
relaxation data discussed below. The behavior of this boundary at small
doping levels is not established in the present work and needs further
investigation. The right-hand edge of this boundary is drawn in agreement
with data discussed in \cite{r10}.

Two samples of YBa$_2$Cu$_3$O$_7$ (YBCO7) made by a solid-state reaction
were used in our studies. The critical temperatures (T$_{c-onset}$) were
measured on a SQUID and were equal to 91.6 K and 91.2 K for the samples \#1
and \#2, respectively. Home-built pulsed coherent NMR/NQR spectrometers were
used for copper NQR measurements.

The transverse Cu(2) nuclear relaxation was studied in the temperature range
4.2-77 K. The decay curves were fitted to a following expression: 
\begin{equation}
M(\tau )=M_0\cdot \exp \left( -\frac{(2\tau )^2}{2T_{2G}^2}\right) \cdot
\exp \left( -\frac{2\tau }{T_{2L}}\right) .
\end{equation}
The first exponent in this expression represents the well-known Gaussian
decay of the magnetization due to indirect Cu(2) nuclei interaction\cite{r17}%
, the second one accounts for a possibility for an additional transverse
relaxation channel. The Redfield contribution to the transverse relaxation
becomes significant only at T$>$60-70 K and was not taken into account. The
measured transverse relaxation rates, T$_{2G}^{-1}$ and T$_{2L}^{-1}$, are
shown in Fig.2. As is seen in Fig.2 the Gaussian contribution to the total
rate does not depend on temperature below T$_c$, and the additional
Lorenzian one exibits one maximum, at T = 35 K, for the sample \#1 and two
maxima, at T = 35 K and T = 47 K, for the sample \#2.

The Cu(2) nuclear spin-lattice relaxation was studied in the temperature
range 4.2-300 K by the saturation-recovery method. The recovery curves were
fitted to an expression: 
\begin{equation}
\frac{M(t)-M_0}{M_0}=\exp \left( -\frac t{T_1}\right) ^N,
\end{equation}
with M$_0$ an equilibrium magnetization, M(t) a magnetization after a
saturating $\pi /2$-pulse, and the parameter N characterizes the relaxation
rate distribution: N=1 for a homogeneous system for which all nuclei have
the same relaxation rate, and N$<$1 for a inhomogeneous system for which the
fluctuating fields causing the relaxation are different for different nuclei.

The temperature dependence of Cu(2) nuclear spin-lattice relaxation rate is
shown in Fig.3. The dependence is the same for both samples between 10 and
300 K. Below T$_c$ the rate decreases rapidly which is known to be due to a
superconducting gap appearing in the excitation spectrum of a
superconductor. The spin-lattice relaxation kinetics gradually becomes
non-exponential below 35-40\ K (N$<$1). No any peculiarities of the
relaxation rate are observed at T = 35 K and 47 K.

One can see in Fig.1 that near optimal doping the phase boundaries are
crossed twice when decreasing temperature along the line 1. Since the phase
boundary depicted by a dashed line in Fig.1 depends strongly on a doping
level, the small changes of it leave only one crossing of the phase boundary
(for example, when the temperature is decreasing along the line 2). This
crossing can be naturally connected with a T$_2^{-1}$ maximum at T$\approx $%
35 K.

The longitudinal nuclear magnetic relaxation is known to be produced by the
fields fluctuating at a NMR frequency and directed perpendicularly to the
quantization axis which coincides with the electric field gradient in the
case of NQR. The transverse relaxation is caused by the low-frequency
fluctuating fields directed along the quantization axis. Since two copper
isotopes, $^{63}$Cu and $^{65}$Cu, have different gyromagnetic ratios and
quadrupolar moments ($^{63}\gamma /^{63}\gamma =0.933$, $^{63}Q/^{65}Q=1.081$%
), the measurements for different isotopes allow to distinguish between the
relaxation caused by fluctuating magnetic or electric fields. Moreover, if
the transverse relaxation is caused by interaction between like nuclear
spins it should be 1.3 times slower for $^{65}$Cu than for $^{63}$Cu due to
bigger concentration (natural abundance) of the latter.

Orbital currents circulating in CuO$_2$ planes create fluctuating magnetic
fields at Cu(2) sites directed along the {\it c}-axis of the crystal. Since
the electric field gradient at Cu(2) sites is also directed along the {\it c}%
-axis, the changes in the spontaneous currents phase will lead to the
critical acceleration predominantely of the transverse nuclear relaxation
leaving the spin-lattice relaxation unchanged. This corresponds to the
experiment fairly well (see Fig.3a). The comparison of the relaxation rates
for the $^{63}$Cu and $^{65}$Cu isotopes confirms the magnetic character of
an additional channel of transverse relaxation at T = 47 K and indicates
that it is not connected to the interaction between like nuclear spins: $%
^{63}T_{2L}^{-1}/^{65}T_{2L}^{-1}=0.89\approx (^{63}\gamma /^{65}\gamma )^2$.

We could not definitely determine the character of additional relaxation
channel at T = 35 K due to insufficient temperature stability during the
long signal aquisition time. According to \cite{r7} the additional
relaxation at 35 K in the underdoped YBCO7 is caused by fluctuations of
electric field gradient at Cu(2) sites. This could be explained if one
assumes that almost nothing happens to the spontaneous currents phase while
crossing the phase boundary along the line 2 in Fig.1, but that an
inhomogeneous superconducting phase with a spacially modulated charge
(U-phase in terms of \cite{r11}) appears below 35 K. The observed changes in
the character of the magnetization recovery is also indicative of the phase
transition into inhomogeneous (incommensurate) state: below 30-40 K the
parameter N in Eq.(2) decreases below 1 (see Fig.3b), i.e., the recovery
curve begins to deviate from an exponent. It is worth to recall Fig.4,5 of
the paper \cite{r19} in this connection, in which the temperature behavior
of the Gaussian Cu(2) spin-spin relaxation is reported to be qualitatively
different above and below T$\approx $30 K in YBa$_2$Cu$_3$O$_{6.98}$ and LaBa%
$_2$Cu$_3$O$_{6.93}$. This fact also can be considered as an indication of a
phase transition at about 30 K.

Strong dependence of the position of the second maximum in T$_2^{-1}$ on the
oxygen deficiency (see Table 1) corresponds fairly well to the strong doping
dependence of the temperature at which the phase of spontaneous currents
disappears. The incline of the line separating the {\it d}-SC and the mixed (%
{\it d}-SC+{\it id}-CDW) phases to the left in Fig.1 indicates the tendency
of expellation of the spontaneous currens phase from the superconducting
phase. This tendency has been already theoretically predicted in \cite{r11}.

Finally, let us give an additional argument in favour of inhomogeneous
superconducting phase. Already at T$>$T$_c$ the chain structures of YBCO7
exhibit the behavior typical for the systems with CDW (see, for example, 
\cite{r20}). Hybridization of the chain and plane energy bands promotes the
inhomogeneous superconductivity also in the planes. It becomes clear in this
connection why there are no anomalies in the transverse nuclear relaxation
rate of copper in La$_{2-x}$Sr$_x$CuO$_4$ at about 30 K \cite{r3}.

The work is supported by the Russian program ''Superconductivity'', under
Grant No.98014-1.

\section{Figure captions}

Fig.1. The schematic phase diagram of YBa$_2$Cu$_3$O$_{7-y}$. As in \cite
{r10}, {\it p} denotes the hole concentration per one Cu(2) position.

Fig.2. Temperature dependence of $^{63}$Cu(2) transverse relaxation rates
for the sample \#1 (a) and \#2 (b). The vertical lines correspond to T=35 K
and 47 K.

Fig.3. Temperature dependence of $^{63}$Cu(2) spin-lattice relaxation rate
(a) and of parameter N (b, see text for details). The vertical lines
correspond to T=35 K, 47 K and 91.4 K.

\newpage

\begin{table}[tbp]
\caption{Positions of maxima of Cu(2) transverse relaxation rate below $T_c$%
. }
\label{t1}
\begin{tabular}{|c|c|c|c|c|}
\hline
Compound & T$_c$,\,K & Max.1,\,K & Max.2,\,K & Reference \\ \hline
YBa$_{2}$Cu$_{3}$O$_{6.98}$ & 92 & 35 & 52 & [5] \\ \hline
YBa$_{2}$Cu$_{3}$O$_{7-y}$ & 90 & 35 &  & [7] \\ \hline
YBa$_{2}$Cu$_{3}$O$_{7-y}$ & 93 & 35 &  & [3] \\ \hline
YBa$_{2}$Cu$_{3}$O$_{6.95}$ & $\approx91$ & 35 & 50 & [9] \\ 
YBa$_{2}$Cu$_{3}$O$_{6.925}$ & $\approx91$ & 35 &  &  \\ \hline
YBa$_{2}$Cu$_{3}$O$_{7-y}$ & 91.6 & 35 & - & Our results \\ 
& 91.2 & 35 & 47 &  \\ \hline
\end{tabular}
\end{table}

\end{document}